# Multi-Analyte, Swab-based Automated Wound Monitor with AI

Madhu Babu Sikha[1], Lalith Appari[1], Gurudatt Nanjanagudu Ganesh[2], Amay Bandodkar[2], Imon Banerjee[1,3]
[1]Department of Radiology, Mayo Clinic Arizona, Phoenix, AZ, USA.
[2]Department of Electrical and Computer Engineering, North Carolina State University, Raleigh, NC, USA.
[3]School of Electrical, Computer and Energy Engineering, Arizona State University, Tempe, AZ, USA.

*Abstract*— **Diabetic foot ulcers (DFUs), a class of chronic wounds, affect ~750,000 individuals every year in the US alone and identifying non-healing DFUs that develop to chronic wounds early can drastically reduce treatment costs and minimize risks of amputation. There is therefore a pressing need for diagnostic tools that can detect non-healing DFUs early. We develop a low cost, multi-analyte 3D printed assays seamlessly integrated on swabs that can identify non-healing DFUs and a Wound Sensor iOS App - an innovative mobile application developed for the controlled acquisition and automated analysis of wound sensor data. By comparing both the original base image (before exposure to the wound) and the wound-exposed image, we developed automated computer vision techniques to compare density changes between the two assay images, which allow us to automatically determine the severity of the wound. The iOS app ensures accurate data collection and presents actionable insights, despite challenges such as variations in camera configurations and ambient conditions. The proposed integrated sensor and iOS app will allow healthcare professionals to monitor wound conditions real-time, track healing progress, and assess critical parameters related to wound care.**

## I. Introduction

Diabetic foot ulcers (DFUs) represent a significant clinical challenge, affecting approximately 750,000 individuals annually in the United States alone. As a class of chronic wounds, DFUs pose high risks of infection, delayed healing, and, in severe cases, lower-limb amputation [1,2]. Early identification of non-healing DFUs—those likely to progress into chronic, treatment-resistant wounds—is critical for improving patient outcomes and reducing healthcare costs [3,4]. However, current clinical assessment methods often rely on subjective visual inspection and lack standardized, quantitative diagnostics, resulting in delayed intervention and inconsistent care.

Algorithms such as ''NERDS'' and ''STONEES'' [5] that rely on visual clinical signs have been developed to detect wound characteristics, however, their accuracy remains controversial, and none have been widely adopted. The lack of tools for quantitative, evidence-based wound diagnosis impedes clinicians from providing wound-specific treatments which has been a major contributor to dismal healing rates for DFUs with ~70% of DFUs remaining unhealed [6], a percentage that hasn't changed in decades.

Recent examples of smart dressings that detect wound biomarkers in a real-time fashion are promising but they face several practical, regulatory, and financial challenges starting from maintaining intimate sensor-wound contact, prone to failure especially when applied to wounds in constricted body parts (sacrum, foot) and expensive. More importantly, the caregiver/patient must be trained in using these dressings [7] and the dressings' reliance on wireless electronics introduces another layer of complexity and point of system failure which can increase patient/caregiver stress levels.

Some studies have explored direct wound imaging for assessment using mobile apps or computer vision, focusing on wound features like size, shape, and color [8,9,10]. These methods face challenges, including the lack of biochemical data for healing, difficulties in capturing images of hard-to-reach areas like the feet, lighting issues, and privacy concerns.

To address this unmet need, we present an integrated, low-cost sensor-based diagnostic solution that combines a novel multi-analyte, 3D-printed assay with a mobile-based imaging and computer vision monitoring system. The assay is embedded on standard swabs and engineered to respond to key wound biomarkers, enabling non-invasive and point-of-care monitoring of wound status. Complementing this hardware, we developed WoundSensor, a dedicated iOS application designed for controlled image acquisition and automated interpretation of assay responses based on computer vision.

## II. Methodology

### A. Details of 3D printed Assays

Assays were fabricated by incorporating the enzymatic swabs on a 3D printed base. The base was 3D printed to accommodate multiple swabs capable of individual colorimetric detection of analytes. For the proof-of-concept, we developed lactate sensing swabs for measurement in wound exudates. To achieve this, we punched out the peroxidase swabs, each with 2 mm diameter and drop-coated them with 120 mg/mL lactate oxidase enzyme. The enzyme was coated 0.5 µL at a time to a final volume of 2 µL. Once the enzyme is dried on the swab surface, 0.5 µL of chitosan (0.5%) was added and the swabs were placed in 4°C for 1 hour. Immediately after, 10 µL of glutaraldehyde was dropped around the swabs to create the fumigated atmosphere and kept at 4°C overnight. The chitosan and glutaraldehyde treatment creates a caged structure on the swab, which avoids the enzyme leaching out during the testing. The as-prepared swabs are then placed in the 3D printed base to fabricate the final assay. The reaction on the assay surface with the introduction of wound exudate proceeds as show in Fig. 1.

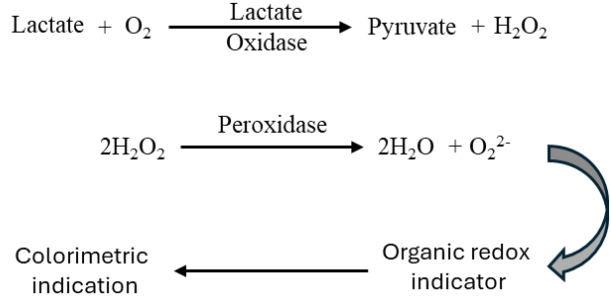

Fig. 1: Reaction mechanism of colorimetric lactate detection.

*B. Mobile application and Image Processing*

We developed an iOS application that follows several key steps to accurately extract and compare colorimetric information from assay images, both before and after exposure, as shown in Fig. 2. The app is configured to control image quality by limiting zoom and adapting to ambient lighting conditions. Additionally, automated image capture is integrated and triggered only when the assay is properly aligned within the view.

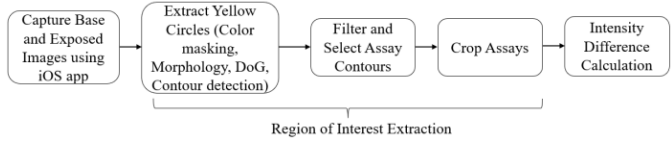

Fig. 2: Proposed iOS application workflow - image processing pipeline.

We captured two RGB images of the assay post: a base image ($I^{base}$) of the post before exposure to the wound and an exposed image ($I^{exposed}$) after interaction with the wound environment. These images serve as the basis for subsequent analysis and finally identifying the intensity changes between the two images.

*Region of Interest (ROI) extraction:*

(i) *Color Space conversion:* The input image $I(x,y,c)$ ($I^{base}$ or $I^{exposed}$) is converted from the RGB color space to the HSV color space because HSV is more suitable for color-based segmentation, as it separates color information from intensity, hence more robust to lighting variations than RGB. Here $c$ represents the channel in the input image, $c \in \{R, G, B\}$.

$$I_{HSV}(x,y,c') = T_{RGB \rightarrow HSV}[I(x,y,c)] \quad (1)$$

where $c' \in \{H, S, V\}$ represents the hue (H), saturation (S) and value (V) channels and $T_{RGB \rightarrow HSV}$ is the transformation function.

(ii) *Yellow color masking:* Given base color of the assays are yellow, a binary mask $M_{yellow}(x,y)$ is created by thresholding the HSV image based on predefined lower ($L_{yellow}$) and upper ($U_{yellow}$) bounds.

$$M_{yellow}(x,y) = \begin{cases} 1, & \text{if } L_{yellow} \leq I_{HSV}(x,y,c') \leq U_{yellow}, \forall c' \\ 0, & \text{otherwise} \end{cases} \quad (2)$$

(iii) Morphological operations: To reduce noise and fill gaps in the segmented regions, morphological opening (dilation) and closing (erosion) operations were applied to this mask.

$$M_{opening}(x,y) = (M_{yellow} \ominus K) \oplus K$$
$$M_{closing}(x,y) = (M_{opening} \oplus K) \ominus K \quad (3)$$

where $\ominus$ is erosion, $\oplus$ is dilation and $K$ is structuring element. The resulting mask was used to extract the yellow-colored pixels from the original image.

$$I_{masked}(x,y,c) = I(x,y,c) \cdot M_{closing}(x,y) \quad (4)$$

(iv) Blob detection using Difference of Gaussians (DoG): The grayscale version $I_{gray}(x,y)$ of the masked image $I_{masked}(x,y,c)$ is used for enhancing blob-like structures corresponding to the four assays in the post using DoG method [11]. The Gaussian blur is applied with two different standard deviations ($\sigma_1$ and $\sigma_2$):

$$G_1(x,y) = I_{gray}(x,y) * \mathcal{G}(x,y,\sigma_1)$$
$$G_2(x,y) = I_{gray}(x,y) * \mathcal{G}(x,y,\sigma_2) \quad (5)$$

where $*$ denotes convolution and $\mathcal{G}(x,y,\sigma) = \frac{1}{2\pi\sigma^2} e^{-\frac{x^2+y^2}{2\sigma^2}}$. Now, the DoG is calculated as,

$$DoG(x,y) = G_1(x,y) - G_2(x,y) \quad (6)$$

Now, apply threshold on the DoG image to obtain binary blobs.

$$B_{blobs}(x,y) = \begin{cases} 1, & \text{if } DoG(x,y) > T_{blob} \\ 0, & \text{otherwise} \end{cases} \quad (7)$$

where $T_{blob}$ is the threshold.

(v) Contour detection and filtering: We find contours of the detected blobs by thresholding the binary mask. The list of contours $\mathcal{C} = contours(B_{blobs}(x,y)) = \{C_1, C_2, ..., C_n\}$, where $C_i$ represents $i^{th}$ contour. We considered only four contours with maximum area which are assumed to correspond to four assays. The filtered list of contours is, $\mathcal{C}' = \{C_1, C_2, C_3, C_4\}$. For each detected contour $C_i$ in both the base and exposed images, the individual assay area was cropped using a mask created from the contour.

$$M_{C_i}(x,y) = \begin{cases} 1, & \text{if } (x,y) \text{ is inside the boundary of } C_i \\ 0, & \text{otherwise} \end{cases} \quad (8)$$

Here $(x,y)$ represents the pixel coordinates in the original uncropped image. Let $(x_i, y_i, w_i, h_i)$ represent the bounding box of contour $C_i$, the cropped image for assay $i$ is calculated as:

$$A_i(x', y', c) = I(x_i + x', y_i + y', c) \cdot M_{C_i}(x_i + x', y_i + y') \quad (9)$$

where $0 \leq x' < w_i$, $0 \leq y' < h_i$; $w_i$ and $h_i$ are the width and height of contour $C_i$. Finally, a larger square region encompassing all four detected assays is cropped from the original image.

*Intensity difference calculation:*
For each corresponding assay in the base and exposed images, the mean intensity of each color channel is calculated, only for the non-zero pixels within the cropped assay region. These pixels are selected with the help of the contour mask.

$$\mu_{i,c}{}^{base} = \frac{\Sigma_{x^i, y^i} A_i{}^{base}(x', y', c) \cdot M_{C_i}(x_i + x', y_i + y')}{\Sigma_{x^i, y^i} M_{C_i}(x_i + x', y_i + y')} \quad (10)$$

where $\mu_{i,c}{}^{base}$ represents the average pixel intensity channel $c$ of $i^{th}$ assay. Similarly, $\mu_{i,c}{}^{exposed}$ is calculated and is used to find the intensity difference of channel $c$ of $i^{th}$ assay as given below:

$$\Delta I_{i,c} = \mu_{i,c}{}^{exposed} - \mu_{i,c}{}^{base} \quad (11)$$

### III. RESULTS AND DISCUSSION

In this section we discuss the results of iOS application workflow. The following parameters are set in the experiment to ensure proper detection of the assays. For yellow color masking, the lower ($L_{yellow}$) and upper ($U_{yellow}$) bounds in HSV color space were set to [20, 150, 150] and [30, 255, 255], respectively. Morphological operations were applied with a rectangular structing element of size $K = 5$. The values of $\sigma_1$ and $\sigma_2$ in blob detection using DoG method are set to 1.5 and 3.0, respectively. Blob detection was performed with a threshold $T_{blob} = 100$.

Fig. 3 shows the screenshot of our iOS app which shows the base image, exposed image and results displayed on the results page. Fig. 4 illustrates the visual and quantitative changes observed in the colorimetric assays before and after exposure to the wound environment. Fig. 4(a) presents an iOS app captured image of the prepared assay post prior to exposure, which was used as the baseline for colorimetric analysis. After exposure to the wound exudate containing lactate, it can be observed from Fig. 4(b) that the yellow colorimetric assays have become darker because of the reaction with the target analyte. To quantify these color changes, Fig. 4(c) displays the mean intensity values for the blue, green, and red color channels for corresponding regions, and the difference in both the base and exposed images calculated automatically using (11). It can be observed that the decrease in mean intensity values for both the blue and green channels in the exposed image compared to the baseline indicates a darkening effect because of the interaction with the wound exudate. A similar trend of values can be observed from Fig. 5 also, which corresponds to another set of base and exposed images.

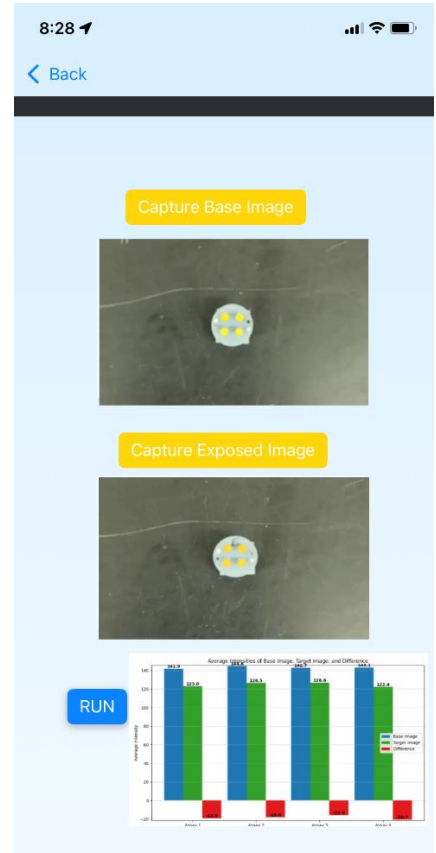

Fig. 3: Screenshot of iOS app results page

*Challenges:*
One significant challenge encountered during the initial stages of this work was the size of the assay and overall post size. With a 1 mm diameter assay on a roughly 1 cm square post, the iOS app's autofocus mechanism struggled to achieve sufficient clarity when using the main camera sensor. When we crop the assay region from the captured images, they resulted in blurry images, which led to inaccurate results. While a macro sensor might be a good solution for capturing closer objects, two main reasons stopped us from using it: first, the color representation might not be accurate, and second, not all user devices have a secondary sensor. These two reasons led us to disable the automatic switching to the secondary sensor within the iOS app. Upon increasing the assay diameter to 2 mm, we observed a noticeable improvement in the cropped assay images and a significant improvement in perceiving the intensity difference between the base and exposed images.

### IV. CONCLUSION

Proposed system integrates 3D printed assay and advanced computer vision algorithms to analyze the density and colorimetric changes in the assay before and after contact with the wound environment. By comparing the baseline (pre-exposure) and post-exposure assay images, the app quantifies wound severity metrics in real-time. The mobile platform experience challenges related to variations in camera hardware and ambient lighting conditions, to maintain consistency across diverse clinical settings. Our team is working to solve the challenges from the computer vision side with tight collaboration with the sensor team.

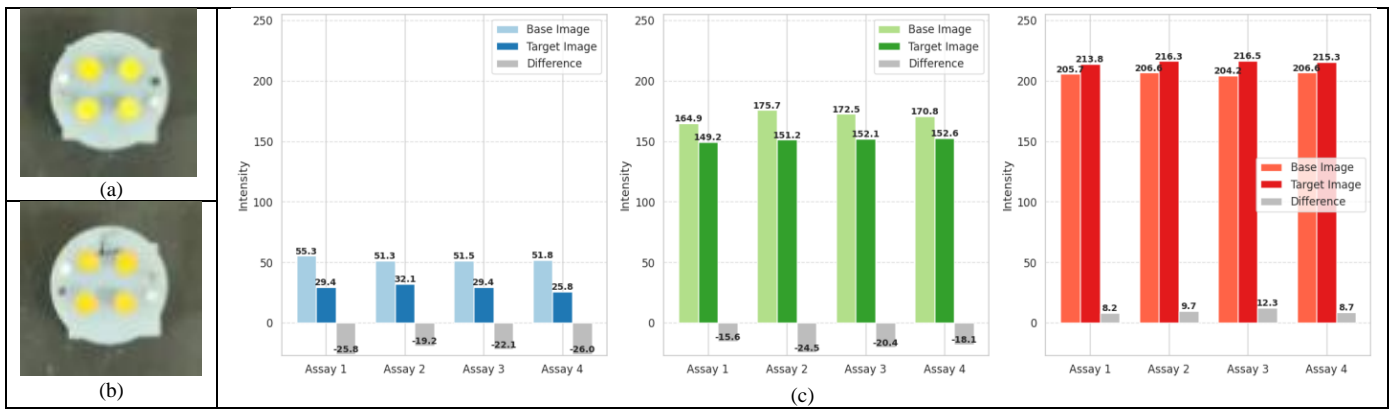

Fig. 4: Results of iOS application workflow for a 2 mm diameter assay post (a) base image (b) exposed image with 30mM lactate concentration and (c) channel-wise intensity values: blue, green and red [left to right]. Each channel has intensity values for base image, exposed image and their difference for four assays.

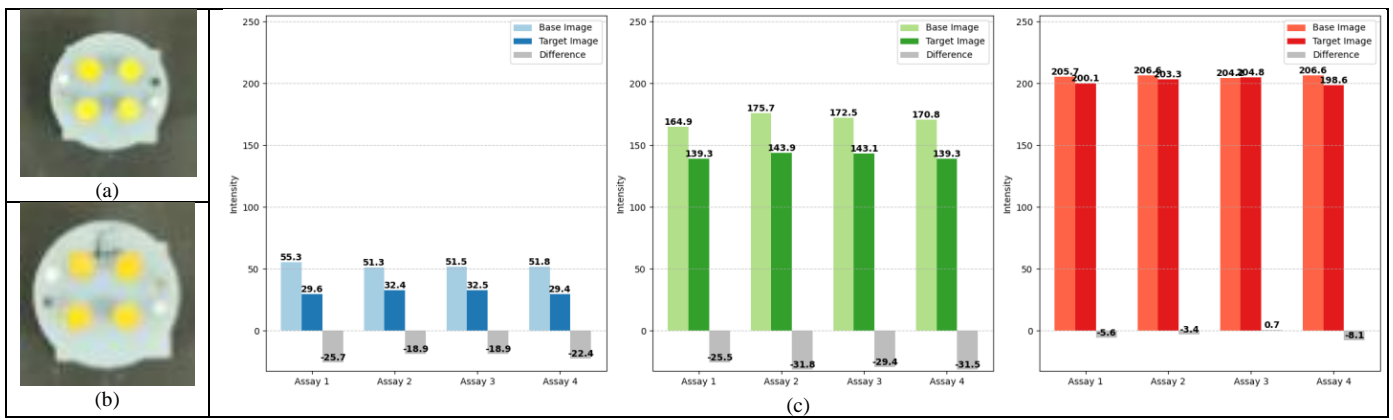

Fig. 5: Results of iOS application workflow for a 2 mm diameter assay post (a) base image (b) exposed image with 30mM lactate concentration and (c) channel-wise intensity values: blue, green and red [left to right]. Each channel has intensity values for base image, exposed image and their difference for four assays.

Once implemented, this integrated wound monitoring solution will enable healthcare professionals to detect non-healing wounds early, track healing trajectories longitudinally, and make data-driven decisions at the point of care. Our approach represents a scalable, user-friendly advancement in chronic wound management, with the potential to improve outcomes and reduce the burden of DFU-related complications.

ACKNOWLEDGMENT

This work is supported by the National Institutes of Health under grant number R01 EB035403-02.

REFERENCES

[1] Pitocco D, Spanu T, Di Leo M, Vitiello R, Rizzi A, Tartaglione L, Fiori B, Caputo S, Tinelli G, Zaccardi F, Flex A, Galli M, Pontecorvi A, Sanguinetti M, "Diabetic foot infections: a comprehensive overview," Eur Rev Med Pharmacol Sci., vol. 23(2 Suppl), pp. 26-37, Apr. 2019.

[2] McDermott K, Fang M, Boulton AJM, Selvin E, Hicks CW, "Etiology, Epidemiology, and Disparities in the Burden of Diabetic Foot Ulcers," Diabetes Care, vol. 46, no. 1, pp. 209-221, Jan. 2023.

[3] Chang M, Nguyen TT, "Strategy for Treatment of Infected Diabetic Foot Ulcers," Acc Chem Res, vol. 54, no. 5, pp. 1080-1093, Mar. 2021.

[4] William J. Jeffcoate, Loretta Vileikyte, Edward J. Boyko, David G. Armstrong, Andrew J.M. Boulton, "Current Challenges and Opportunities in the Prevention and Management of Diabetic Foot Ulcers," Diabetes Care, vol. 41, no. 4, pp. 645–652, Apr. 2018.

[5] Woo KY, "The use of antimicrobial dressings in chronic wounds: NERDS and STONEES principles," Surg Technol Int., vol. 20, pp. 73-82, Oct. 2010.

[6] Armstrong DG, Tan TW, Boulton AJM, Bus SA, "Diabetic Foot Ulcers: A Review," JAMA, vol. 330, no. 1, pp. 62-75, Jul. 2023.

[7] O'Callaghan, Suzanne, Paul Galvin, Conor O'Mahony, Zena Moore, and Rosemarie Derwin, "'Smart' wound dressings for advanced wound care: a review." Journal of wound care, vol. 29, no. 7, pp. 394-406, Jul. 2020.

[8] Ploderer B, Brown R, Seng LSD, Lazzarini PA, van Netten JJ, "Promoting Self-Care of Diabetic Foot Ulcers Through a Mobile Phone App: User-Centered Design and Evaluation," JMIR Diabetes, vol. 4, no. 4, Feb. 2018.

[9] Anthony CA, Femino JE, Miller AC, Polgreen LA, Rojas EO, Francis SL, Segre AM, Polgreen PM, "Diabetic Foot Surveillance Using Mobile Phones and Automated Software Messaging, a Randomized Observational Trial," Iowa Orthop J., vol. 40, no. 1, pp. 35-42, 2020.

[10] Swerdlow M, Shin L, D'Huyvetter K, Mack WJ, Armstrong DG, "Initial Clinical Experience with a Simple, Home System for Early Detection and Monitoring of Diabetic Foot Ulcers: The Foot Selfie," Journal of Diabetes Science and Technology, vol. 17, no. 1, pp. 79-88. Oct. 2021.

[11] Rafael C. Gonzalez, Richard E. Woods, Steven L. Eddins, "Digital Image Processing using MATLAB", 3rd ed., Gatesmark publishing, Jan. 2020.